\documentclass[prd,aps,10pt,twocolumn,floats,nofootinbib,eqsecnum,showpacs]{revtex4-1}

\usepackage{amssymb,amsmath}
\usepackage{epsfig}
\pdfoutput=1
\usepackage{amssymb,amsmath}
\usepackage{epsfig}
\usepackage{xcolor}
\usepackage[utf8]{inputenc}
\usepackage[normalem]{ulem}

\DeclareMathOperator\erf{erf}

\DeclareMathOperator\erfi{erfi}

\newcommand{\be}{\begin{equation}}
\newcommand{\ee}{\end{equation}}
\newcommand{\ba}{\begin{eqnarray}}
\newcommand{\ea}{\end{eqnarray}}
\newcommand{\beg}{\begin{gather*}}
\newcommand{\eng}{\end{gather*}}
\newcommand{\hh}{,\hspace{0.5cm}}

\newcommand{\eq}[1]{(\ref{#1})}

\newcommand{\n}[1]{\label{#1}}

\newcommand{\ins}[1]{{\mbox{\tiny #1}}}

\def\XXint#1#2#3{{\setbox0=\hbox{$#1{#2#3}{\int}$ }
\vcenter{\hbox{$#2#3$ }}\kern-.6\wd0}}

\usepackage[makeroom]{cancel}
\usepackage[caption=false]{subfig}
\usepackage[colorlinks=true,
            citecolor=green,
            linkcolor=red,
            filecolor=cyan,
            urlcolor=magenta,
            backref=false]{hyperref}

\newcommand{\dd}{\mbox{d}}

\begin{document}

\title{Quantum scattering on a delta potential in ghost-free theory}

\author{Jens Boos}
\email{boos@ualberta.ca}
\affiliation{Theoretical Physics Institute, University of Alberta, Edmonton, Alberta, Canada T6G 2E1}

\author{Valeri P. Frolov}
\email{vfrolov@ualberta.ca}
\affiliation{Theoretical Physics Institute, University of Alberta, Edmonton, Alberta, Canada T6G 2E1}
\author{Andrei Zelnikov}
\email{zelnikov@ualberta.ca}
\affiliation{Theoretical Physics Institute, University of Alberta, Edmonton, Alberta, Canada T6G 2E1}

\date{\today}

\begin{abstract}
 We discuss the quantum-mechanical scattering of a massless scalar field on a $\delta$-potential in a ghost-free theory and obtain analytic solutions for the scattering coefficients. Due to the non-locality of the ghost-free theory the transmission coefficient tends to unity for frequencies much larger than the inverse scale of non-locality, even for infinitely strong potentials. At the same time there exists a critical strength of the $\delta$-potential barrier below which there is always a frequency that is totally reflected. These scattering properties in ghost-free theories are quite generic and distinguish them from local field theories. Moreover, we study quasi-normal states that are present for the $\delta$-potential well. In the limit of vanishing non-locality, we recover the standard results of local field theory.
\end{abstract}

\pacs{03.65.Ge,03.65.Nk,03.65.Pm \hfill Alberta-Thy-05-18\\[\baselineskip]Keywords: Lippmann--Schwinger equation; non-locality; quasi-normal modes; scattering coefficients}

\maketitle


\section{Introduction}

Exactly solvable problems have played an important role in quantum mechanics, especially at the early stage of its development \cite{Dirac:1947,Landau:1965,Tannoudji:1978,Sakurai:1994,Shankar:1994,Weinberg:1995,Griffiths:1995}. Such problems typically give analytical results for scattering amplitudes and the energies of bound states, and they allow one to analyze non-perturbative aspects of these problems, for example non-analyticity.
A well known problem, both in the non-relativistic and the relativistic case, is the quantum-mechanical scattering of a particle on a $\delta$-like potential. In this letter we consider the same problem in a slightly different setup. Namely, we consider the scattering of a particle on a $\delta$-like potential in the framework of a so-called ghost-free scalar field theory, which belongs to a class of non-local generalizations of relativistic field theories that have recently been studied quite extensively \cite{Modesto:2010uh,Biswas:2011ar,Frolov:2015bia,Modesto:2017ycz,Koshelev:2018hpt}.

A simple example of such a generalization is the field theory of a free scalar massless field, wherein the $1/\Box$ propagator is changed to $f(\ell^2\Box)/\Box$. If $f(z)$ is an exponent of an entire function,  the propagator does not have extra new poles: then, there are no ghost degrees of freedom and thereby unitarity is preserved. An important property of non-local ghost-free field theories is that non-locality becomes important only for off-shell processes, in the presence of external sources or interactions \cite{Buoninfante:2018mre}. All on-shell effects are immune to the ghost-free modifications of the theories. There exists an exhaustive set of publications on this subject, with the main focus on the application of ghost-free gravity to the long-standing problems of General Relativity (cosmological singularities as well as black hole singularities) \cite{Biswas:2010zk,Calcagni:2013vra,Hossenfelder:2009fc,Modesto:2010uh,Zhang:2014bea,Frolov:2015bta,Frolov:2015bia,Conroy:2015wfa,Li:2015bqa,Calcagni:2017sov,Cornell:2017irh,Kajuri:2017jmy,Boos:2018bxf,Koshelev:2018hpt,Buoninfante:2018rlq,Buoninfante:2018stt}. In this letter we discuss a rather simple problem related to the linear regime of ghost-free theory: scattering and resonant states of the ghost-free scalar massless particle in the presence of a $\delta$-like potential. We shall demonstrate that such a problem is also exactly solvable, obtain the reflection and transmission coefficients in closed form, and describe how the energy of the resonance is modified. Certainly, in the limit when the length scale of non-locality $\ell$ tends to zero we shall reproduce the standard results of local quantum mechanics. However, for a finite value of $\ell$ there are quite interesting and unexpected features (at least for us) in the scattering amplitude.

\section{Scalar ghost-free theory}
We start with a non-local modification of the scalar field equation
\be
[{\cal D}- V(X)]\Phi (X) =0 \, .
\ee
Here, ${\cal D}$ is a function of the box operator $\Box$ and $V(X)$ is a local potential term where $X$ are the Cartesian coordinates in a flat spacetime. In a standard local theory one has ${\cal D}=\Box$.

For concreteness let us consider the two-dimensional case. We denote the Cartesian coordinates in the flat spacetime by $X=(t,x)$, and assume that the potential $V$ does not depend on time $t$. Using the Fourier transform, we write $\Phi$ in the form
\be
\Phi(t,x)=\int \limits_{-\infty}^{\infty} \frac{\dd\omega}{2\pi} e^{-i \omega t} \varphi_{\omega} (x)\, .
\ee
 Then, the d'Alembert operator reduces to
\be\label{DE}
\Box\to {\dd^2\over \dd x^2}+E \, , \quad E=\omega^2 \, .
\ee
In what follows, we shall denote by ${\cal D}_E$ the operator ${\cal D}$ to which the substitution \eqref{DE} has been applied.
In the case of a $\delta$-potential, $V=\lambda \delta(x)$, the mode functions $\varphi_{\omega}$ satisfy the following time-independent equation\footnote{In order to simplify the notation, from now on we shall omit the subscript $\omega$ of $\varphi$.}
\be\label{GF}
\left[{\cal D}_E-\lambda\delta(x)\right] \varphi=0 \, .
\ee
The potential describes a $\delta$-potential barrier for $\lambda>0$ and a $\delta$-potential well for  $\lambda<0$.

Our goal is to study a ``ghost-free version'' of this equation. For simplicity, we consider a so-called $\mathrm{GF_1}$ version  of the theory,  where the operator ${\cal D}$ has the form \cite{Frolov:2016xhq}
\be
{\cal D}=\exp{(-\ell^2\Box)} \Box\, .
\ee
In this expression, $\ell$ is the length scale where the non-locality becomes important.

\section{Continuous spectrum. Scattering amplitudes}

\subsection{Local theory}

Before describing solutions of the non-local equation \eqref{GF}, let us first derive the exact solution in the local limit. Note that in the limit $\ell \rightarrow 0$ one has $\mathcal{D} = \Box$, and \eqref{GF} reduces to a time-independent Schr\"odinger-type equation,
 \be \n{VEQ}
\left[{\dd^2\over \dd x^2}+E-\lambda\delta(x)\right]\varphi=0\, .
\ee
It is well known that for both signs of $\lambda$ the energy spectrum of such a problem contains a doubly degenerate continuous part, $E>0$. For negative $\lambda$ there also exists a single bound state with $E<0$ \cite{Griffiths:1995}. In quantum field theory these states correspond to quasi-normal or ringing modes. In this section we consider only the continuous spectrum, and we show that in the $\mathrm{GF_1}$ generalization there also exists a continuous part of the spectrum with $E>0$. However, the scattering amplitudes are modified by the non-locality. We shall turn to the bound states and quasi-normal modes in $\mathrm{GF_1}$ further below.

Suppose we know a solution of the free equation (that is, if $V=0$) of (\ref{VEQ}), which we denote by $\varphi_0$ and let $G_0(x,x')$ be the Green function of the free equation,
\be
\left[{\dd^2\over \dd x^2}+E\right] G_0(x,x')=-\delta(x-x')\, .
\ee
Then a solution of \eqref{VEQ} satisfies the following Lippmann--Schwinger equation \cite{Lippmann:1950zz}:
\be
\label{eq:lippmann-schwinger-solution}
\varphi(x)=\varphi_0(x) - \int\limits_{-\infty}^{\infty} \dd x' G_0(x,x') V(x') \varphi(x')\, .
\ee
The  local free Green function $G_0$ is symmetric, $G_0(x,x')=G_0(x',x)$. In the absence of the potential $V$ the wave equation is translationally invariant, such that the Green function depends only on $z=x-x'$,
\be
G_0(x,x') \equiv {\cal G}_0(z) \, , \quad  {\cal G}_0(z)= {\cal G}_0(-z) \, .
\ee
In the context of scattering theory $\mathcal{G}_0(z)$ must be chosen such that in the time-dependent problem a wave packet constructed with the help of the Green function corresponds to an out-going wave.
Such a Green function is
\be\begin{split}\label{G0}
{\cal G}_0(z)&=\int\limits_{-\infty}^{\infty}{\dd k\over 2\pi}{e^{ikz}\over k^2-\Omega^2-i\epsilon}={i\over 2\Omega} e^{i\Omega |z|} \, .
\end{split}\ee
Here and in what follows we use the notation
\be
\Omega \equiv \sqrt{E} > 0 \, .
\ee
Note that instead of using the properties of time-dependent solutions constructed with the help of the Green function, one can specify a Green function by its analytical properties. In this case one requires that the pole in the complex $E$ plane for the Fourier transform of this Green function is located at $E+i\epsilon$, where $\epsilon$ is a small positive constant. For $E>0$ one has $\sqrt{E+i\epsilon}=\Omega+i\epsilon$. For such a choice of the pole in the integral \eqref{G0} the Green function is decreasing at infinity $|x|\to \infty$. This property allows one to uniquely specify the required Green function.

As is well known, for the $\delta$-like potential the Lippmann--Schwinger equation is exactly solvable. If ${1+\lambda {\cal G}_0(0)}\neq 0$ then the solution reads
\begin{align}
\label{vvv}
\varphi(x) = \varphi_0(x) - \lambda \frac{\varphi_0(0)}{1+\lambda {\cal G}_0(0)} {\cal G}_0(x) \, .
\end{align}
Denoting $\gamma={\Omega/\lambda}$, Eq.~\eqref{vvv} then implies
\be\n{locsol}
\varphi(x) = \varphi_0(x) - \frac{\varphi_0(0)}{1-2i\gamma}\exp(i\Omega |x|)\, .
\ee

To obtain the scattering coefficients let us choose $\varphi_0=\exp( i \Omega x)$. Then, the corresponding time-dependent mode $\exp[-i \Omega (t- x)]$ describes a quantum moving in the positive direction of $x$. Since both the potential and the equation of motion are symmetric under $x \rightarrow -x$, we shall restrict ourselves to the right-moving mode without loss of generality. Thus we have
\begin{align}
\n{ASS}
\varphi(x) &= e^{i \Omega x} - {1\over 1-2i\gamma} e^{i\Omega |x|} \\
\n{pvv}
&=\left\{
\begin{array}{cl}
 {-2i\gamma\over 1-2i\gamma} e^{i\Omega x}  & \mbox{for $x>0$}\, , \\
e^{i\Omega x}-{1\over 1-2i\gamma} e^{-i\Omega x} &\mbox{for $ x<0$}\,  .
\end{array}
\right.
\end{align}
This solution has a well known interpretation: the mode $e^{i\Omega x}$ with unit amplitude propagates from  left to right. It meets the $\delta$-like potential where it is partly reflected and it partly penetrates through it. The corresponding transmission coefficient $t$ and reflection coefficient $r$ are
\be\n{rtsc}
t= -{2i\gamma\over 1-2i\gamma} \, \hh r=-{1\over 1-2i\gamma} \, .
\ee
Then, we can define $T$ and $R$ such that
\begin{align}\label{TR}
T = |t|^2 = \frac{4\gamma^2}{1+4\gamma^2} \, , \quad R = |r|^2 = \frac{1}{1+4\gamma^2} \, .
\end{align}

\subsection{Non-local ghost-free theory}

For the calculation of the scattering coefficients in the ghost-free theory one needs to find a solution of Eq.~\eqref{GF}.
We shall use the same approach as described in the context of local quantum mechanics by employing a representation similar to \eqref{vvv} for a solution of the corresponding Lippmann--Schwinger equation. The only change consists of replacing the local Green function $G_0(x,x')$ with the Green function $G(x,x')$ of the non-local equation without the potential, which satisfies
\be
{\cal D}_E G(x,x')=-\delta(x-x')\, .
\ee
In the case of $\mathrm{GF_1}$ theory the operator ${\cal D}_E$ reads \cite{Frolov:2016xhq}
\be
{\cal D}_E=e^{-\ell^2(\partial^2_x+{E})}(\partial^2_x+{E})\hh
E=\Omega^2\, .
\ee
This operator is translationally invariant, such that its Green function $G$ depends only on $z=x-x'$, just like in the local theory, $G(x,x')={\cal G}(z)$.
Using the Fourier representation for this Green function one obtains
\be\label{calG}
{\cal G}(z)=\int\limits_{-\infty}^{\infty}{\dd k\over 2\pi}e^{ikz}{e^{-\ell^2(k^2-{\Omega^2})}\over k^2-{\Omega^2}-i\epsilon} \, .
\ee
In the above, we added $i\epsilon$ to $\Omega^2$ in accordance with the standard prescription for the description of the scattering problem. The integral \eq{calG} can be easily computed by rewriting it as the sum
\be\label{calGG}
{\cal G}(z)={\cal G}_0(z)+\Delta{\cal G}(z) \, .
\ee
The first term is exactly the Green function \eq{G0} of the local theory. The integrand of the second term is regular. Therefore, the integral is well defined and does not require any $i\epsilon$ prescription:
\be\begin{split}\label{calGG0}
\Delta{\cal G}(z)&=\int\limits_{-\infty}^{\infty}{\dd k\over 2\pi}e^{ikz}{e^{-\ell^2(k^2-\Omega^2)}-1\over k^2-\Omega^2}\\
&=-\int\limits_0^{\ell^2}\dd s \int\limits_{-\infty}^{\infty}{\dd k\over 2\pi}e^{ikz}e^{-s(k^2-\Omega^2)} \, .
\end{split}\ee
By taking the Gaussian integral first and then the integral over the parameter $s$ we arrive at the final result
\be\n{GFGF}
{\cal G}(z) = {i\over 4\Omega}\left[ e^{i\Omega z} Y(z)+ e^{-i\Omega z} Y(-z)\right] \, ,
\ee
where
\be
Y(z)=1+\erf\left(i\alpha+{z\over 2\ell}\right) \hh \alpha=\Omega\ell \, .
\ee
The above Green function has the symmetry ${\cal G}(z)={\cal G}(-z)$ and is expressed in terms of the error function $\erf(x)$ (see, e.g., \cite{Olver:2010}).
The function $Y(z)$ has the following asymptotics:
\be \n{pvv1}
Y(z)\sim\left\{
\begin{array}{rl}
2-{2\ell\over \sqrt{\pi}z}\exp\left[-\left(i\alpha+{z\over 2\ell}\right)^2\right] & \mbox{for  $z\to\infty$ }\, , \\[10pt]
-{2\ell\over \sqrt{\pi}z}\exp\left[-\left(i\alpha+{z\over 2\ell}\right)^2\right] & \mbox{for  $z\to-\infty$ }\,  .
\end{array}
\right.
\ee
Using these relations it is easy to check that the constructed Green function (\ref{GFGF}) has the same property as $\mathcal{G}_0(z)$: for a small shift $\Omega \to \Omega+i\epsilon$ into the complex plane it becomes a decreasing function for $|z|\to\infty$. Also,
\be
{\cal G}(0)={i\over 2\Omega} Y(0)= {i\over 2\Omega}[1+\erf(i\alpha)]\, .
\ee

To calculate the scattering coefficients (under the condition that $1+\lambda\mathcal{G}(0)\not=0$) we write the general solution of the Lippmann--Schwinger equation in the form
\be\label{Lip}
\varphi(x) = \varphi_0(x) - \lambda \frac{\varphi_0(0)}{1+\lambda {\cal G}(0)} {\cal G}(x) \, .
\ee
If we choose ${\varphi_0=\exp(i \Omega x)}$ the solution becomes
\be\n{SOLUTION}
\varphi(x)= e^{i\Omega x}-{\lambda {\cal G}(x) \over 1+{i\lambda\over 2\Omega}[1+\erf(i\alpha)]} \, .
\ee
Since for large $|z|$ the Green function ${\cal G}$ asymptotically coincides with $\mathcal{G}_0(z)$, one can use the relation (\ref{ASS}) in the asymptotic domain with the only change
\be
\label{eq:gamma-nonlocal}
\gamma\to \tilde{\gamma}= \frac{\Omega}{\lambda} -{1\over 2} \erfi(\alpha) \, .
\ee
Let us point out that the imaginary error function $\erfi(x)\equiv-i\erf(ix)$ is real-valued for any real argument $x$. It has the following asymptotics (see, e.g., \cite{Olver:2010}):
\begin{align}
\label{eq:erfi-asymptotics}
\erfi(x) \sim \left\{ \begin{array}{rl} \frac{2x}{\sqrt{\pi}} & \text{ for } x\rightarrow 0 \, , \\[10pt]
\frac{e^{x^2}}{\sqrt{\pi}x} & \text{ for } x\rightarrow \infty \, .
\end{array} \right.
\end{align}
After the change \eqref{eq:gamma-nonlocal} the relations \eqref{rtsc} give the expressions for the reflection and transmission coefficients for the scattering of a quantum on a $\delta$-like potential in the ghost-free $\mathrm{GF_1}$ theory.


\begin{figure*}[!htb]%
    \centering
    \subfloat{{\includegraphics[width=0.47\textwidth]{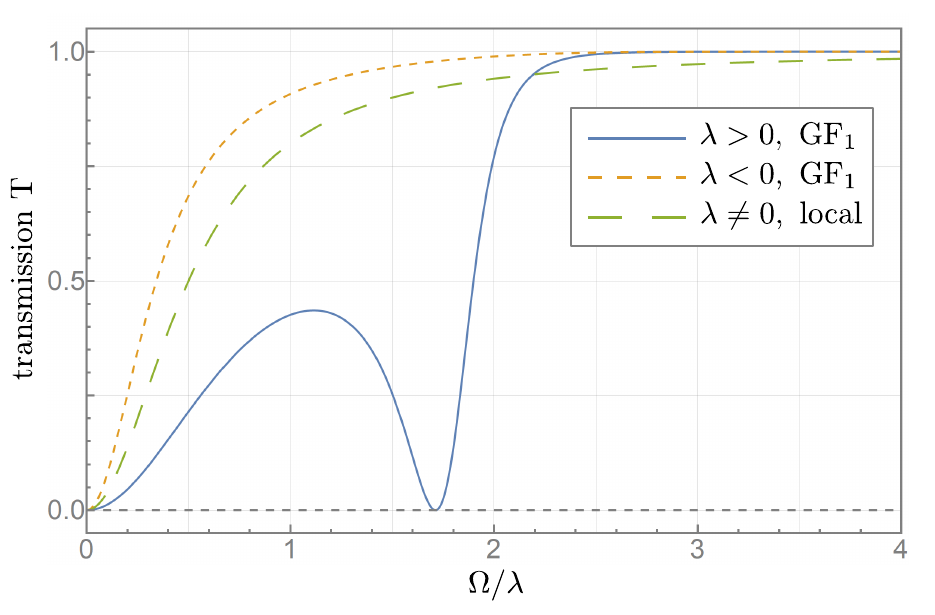} }}%
    \qquad
    \subfloat{{\includegraphics[width=0.47\textwidth]{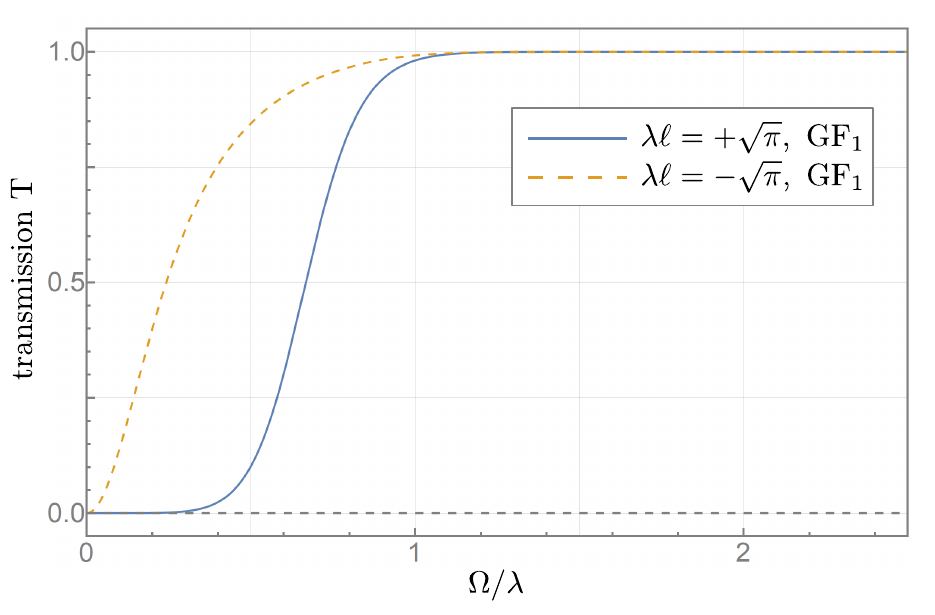} }}

    \caption{These plots show the transmission coefficient $T$ as a function of the dimensionless frequency $\alpha = \Omega/\lambda$. Left plot: In case $0 < \Lambda < \sqrt{\pi}$, the transmission coefficient vanishes for a finite $\alpha=\alpha_\star$. For a $\delta$-potential well ($\Lambda<0$) there is no such minimum in the transmission. In the local case, the transmission coefficients of both the barrier and the well coincide and do not vanish anywhere except at zero frequency. Right plot: For the critical value $\Lambda = \sqrt{\pi}$, all frequencies $\alpha \lesssim 1$ are strongly suppressed. For the $\delta$-potential well with $\Lambda = -\sqrt{\pi}$, nothing special happens.}
    \label{fig:transmission}
\end{figure*}

\section{Properties of scattering coefficients}
To study the properties of the scattering coefficients it is useful to introduce another dimensionless quantity
\be
\Lambda = \lambda\ell \, .
\ee
Using the dimensionless quantities $\alpha$ and $\Lambda$ one obtains
\be
\tilde{\gamma} = {\alpha\over\Lambda}-{1\over 2}\erfi(\alpha) \, .
\ee

The left plot in Fig.~\ref{fig:transmission} shows the transmission coefficient $T$, Eq.~\eqref{TR}, for both positive and negative $\lambda$ in the $\mathrm{GF_1}$ theory, as well as in the local theory. In the non-local case for $\lambda>0$ the transmission vanishes at some finite $\Omega = \Omega_\star$, or, equivalently, $\alpha=\alpha_\star$.
To understand this feature, note that the condition of vanishing transmission coefficient $T=0$ corresponds to $\tilde{\gamma}=0$ and amounts to
\begin{align}
\label{eq:omega-critical}
\frac{2\alpha}{\Lambda} = \erfi(\alpha) \approx \frac{2\alpha}{\sqrt{\pi}} \text{ for } \alpha \ll 1 \, .
\end{align}
Non-trivial solutions $\alpha\not=0$ exist provided
\begin{align}
0<\Lambda < \sqrt{\pi} \, .
\end{align}
For fixed positive $\Lambda$ in this range there is always some finite $\alpha=\alpha_\star$ at which one has total reflection.

For a critical value of hight of the barrier, $\Lambda = \sqrt{\pi}$, the transmission of all modes with $\alpha\lesssim1$ is strongly suppressed, but for higher frequencies $\alpha\gtrsim1$ the barrier becomes transparent; see the right plot in Fig.~\ref{fig:transmission}.

\begin{figure}[!htb]%
    \centering
    \includegraphics[width=0.47\textwidth]{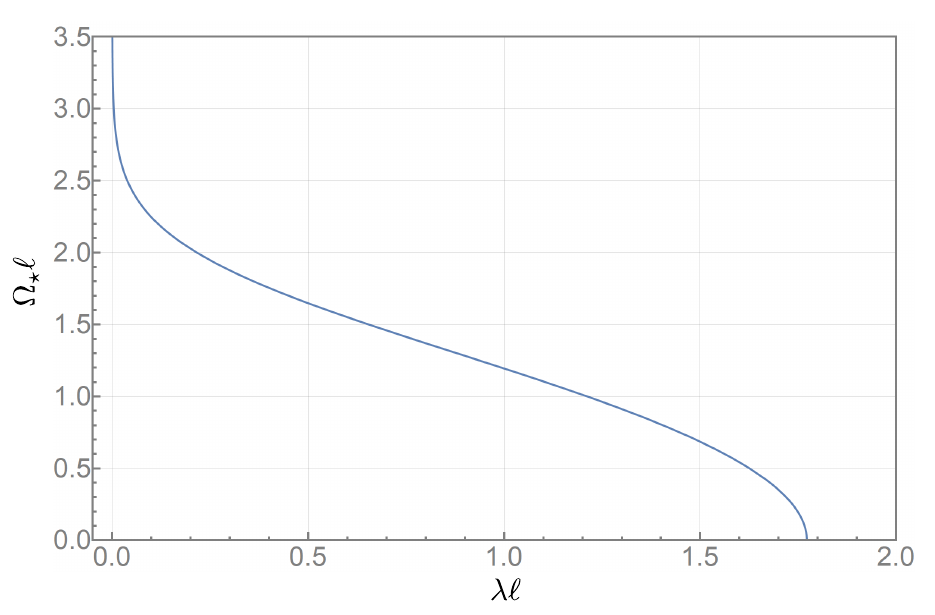}
    \caption{ This plot show the dimensionless frequency $\alpha_\star = \Omega_\star\ell$ as a function of  $\Lambda = \lambda\ell$. For $\Lambda\to 0$ the frequency diverges, but for $\Lambda \approx \mathcal{O}(1)$, the parameter $\alpha_\star$ is finite.  If $\Lambda$ exceeds $\sqrt{\pi}$ the frequency vanishes.}
    \label{fig:critical-omega}
\end{figure}

The condition $\tilde{\gamma}=0$ gives a transcendental equation relating $\alpha_\star$ and $\Lambda>0$; see Fig.~\ref{fig:critical-omega} for a plot of this relation. Considering Eq.~\eqref{eq:omega-critical} in the limit $(\alpha_\star-\alpha) \rightarrow 0$ one obtains the scaling relation
\begin{align}
(\alpha_\star-\alpha) = \sqrt{3}(1 - \Lambda/\sqrt{\pi})^{1/2} \, .
\end{align}

Let us now describe several interesting properties of the non-local scattering coefficients for a fixed, positive value of $\ell$:
\begin{itemize}
\item[(i)] In the limit of an infinitely high $\delta$-potential barrier ($\Lambda\to\infty$) or a $\delta$-potential well ($\Lambda\to-\infty$) the transmission remains finite, $T>0$. This is a distinct property of the non-local ghost-free theory as compared to the local theory.
\item[(ii)]\label{ii}In the case of a $\delta$-potential barrier $0<\Lambda< \sqrt{\pi}$ there is always a frequency $\alpha=\alpha_\star$ for which $\tilde{\gamma}=0$. Hence, this mode is totally reflected, i.e.,\ $T=0$ and $R=1$.
\item[(iii)] In the non-local ghost-free theory, the reflection from the $\delta$-potential depends on the sign of $\lambda$, while in the local theory it does not depend on it.
\end{itemize}

The property that non-local effects enhance the transparency of the $\delta$-potential at high frequencies is not too surprising, because ghost-free deformations of the field theory typically regularize singularities and high frequency behavior. Property (i) is less evident: the potential remains transparent for finite $\alpha$ even in the limit $|\Lambda|\to\infty$ of a very strong potential barrier or potential well.

Property (ii) was unexpected: the $\delta$-potential barrier becomes totally reflective at a rather large frequency $\alpha_\star$ for moderately low potential barriers $0<\Lambda < \sqrt{\pi}$. This property is quite robust and survives for a large variety of ghost-free theories, because for a generic ghost-free theory the parameter $\tilde{\gamma}$ can be read off from the Lippmann--Schwinger equation \eq{Lip}. Expressed in terms of the difference of the ghost-free and local Green functions $\Delta{\cal G}(z)$ it can be written as\footnote{Note that from the definition \eq{calGG0}, as applied to a general ghost-free theory, it follows that $\Delta{\cal G}(0)$ is real.}
\be
\tilde{\gamma}=\Omega\left({1\over\lambda}+\Delta{\cal G}(0)\right).
\ee
The condition of total reflection $\tilde{\gamma}=0$ reduces to
\be
\lambda=-{1\over \Delta{\cal G}(0)}.
\ee
This condition defines the frequency $\Omega_\star$ of the totally reflected mode and can be satisfied for some range of $\lambda$ in a generic ghost-free theory.

We need to point out that a complete, satisfactory physical interpretation of the resonant reflection for this mode is not quite clear. However, the following comment may be useful in this connection. Using the asymptotics \eqref{pvv1} one can present the solution \eqref{SOLUTION} for  $x\to \infty$ as
\be
\varphi(x)=e^{i\Omega x}-B e^{i\Omega |x|}\, ,
\ee
where
\be
B=1-i\left[ {2\alpha\over \Lambda}-\erfi(\alpha)\right].
\ee
One can roughly describe this scattering problem as follows: The interaction of the running wave $e^{i\Omega x}$ with the barrier generates a ``ringing mode'' with a maximum near $x=0$. The decay of this ringing mode creates two running modes moving away from the position of the barrier. Their asymptotic amplitude is $B$. The local and non-local cases differ: both $\varphi(0)$ and the coefficient $B$ are different. In the non-local case for the special value $\alpha=\alpha_\ast$ one has $B=1$. As a result of the interference, at large $x\gg 0$ the ringing mode exactly cancels the original running mode. However, the field $\varphi$ partially penetrates the barrier and does not vanish in a narrow layer behind it. The predicted total reflection at the frequency $\alpha=\alpha_\ast$  is somewhat similar to the effect of high-reflection coatings in optics, where enhanced reflection at a specific frequency is achieved by coating a surface with layers of material with a special profile of an index of refraction. In that sense one may say that ghost-free theory provides a non-local high-reflection coating for the $\delta$-potential.
It should be emphasized that because a typical wavelength of the resonant mode is of the order of the non-locality scale $\ell$ one may expect that an interpretation in terms of traditional local physics fails \cite{Buoninfante:2018mre} and non-locality brings about new physics.


\section{Bound states and quasi-normal modes}
Till now we have mainly focused on the continuous part of the energy spectrum. Let us now discuss states corresponding to discrete components of the energy spectrum. In both local field theory as well as non-local $\mathrm{GF_1}$ theory such discrete levels exist only for $\lambda<0$. In the local theory there is exactly one bound state with negative $E=\Omega^2<0$ \cite{Griffiths:1995}. A normalizable solution with such a (complex) quasi-normal frequency $\Omega$ is often called a quasi-normal or ringing mode (for a discussion of such solutions see, e.g., the nice review \cite{Boonserm:2010px}). In a general case, quasi-normal modes are determined by the poles of the transmission amplitude. We shall demonstrate that for a small value of the dimensionless parameter $|\Lambda|$, a similar unique normalizable solution with pure imaginary frequency exists also in the $\mathrm{GF_1}$ theory. However, for large values of the non-locality there may be a discrete set of solutions with complex frequencies.

In order to find the solutions for quasi-normal modes one can use the Lippmann--Schwinger equation \eq{Lip}. It is sufficient put there $\varphi_0(x)=0$. Then, for the ghost-free theory with attractive $\delta$-like potential ($\lambda<0$), it gives
\be\label{QNM}
\varphi(x) = -\lambda\, {\cal G}(x)\varphi(0) \, .
\ee
This equation has a non-trivial solution only when the condition
\be\label{QNf}
1+\lambda {\cal G}(0)=0
\ee
is satisfied. Here, ${\cal G}(0)$ depends on $\Omega$, but $\Omega$ is to be considered as a complex variable. The solutions $\Omega_\ins{QN}$ of this equation correspond to the poles of the transmission coefficient for a generic ghost-free theory.

One can see that the solution is proportional to the ghost-free Green function
\be
\varphi(x) = c\, {\cal G}(x) \, .
\ee
Quasi-normal modes are assumed to be normalizable, i.e., the factor $c$  is finite. It means that among all the solutions we have to choose those that decrease at infinity.

In the local case ${\cal G}_0(0)=i/(2\Omega)$ and we reproduce the classical result for the unique quasi-normal frequency
\be
\Omega_\ins{QN}=-i\,{\lambda\over 2} \, , \quad \lambda<0 \, .
\ee
The corresponding quasi-normal mode decreases at infinity and is normalizable.

In the non-local theory there may be a set of quasi-normal modes with complex frequencies, depending on the value of $\lambda$.
However, there always exists only one solution with purely imaginary frequencies $\alpha=i\kappa$, where $\kappa>0$. For this case  in $\mathrm{GF_1}$ theory the relation \eqref{QNf} takes the form
\be\label{QN1}
1+{\Lambda\over 2\kappa}\left[ 1-\erf(\kappa)\right] = 0\, \hh  \Lambda<0 \, .
\ee
For $\Lambda<0$ and $\kappa \in \mathbb{R}$ this equation has a unique solution. Also, numerical investigations have shown that for
\be \n{INEC}
|\Lambda |<  2.282365
\ee
this is the only solution of \eqref{QN1} with complex $\kappa$. However, if the inequality (\ref{INEC}) is violated, additional complex solutions appear besides this pure imaginary solution.

\section{Discussion}

Let us discuss the results presented in this letter. Our starting point was a two-dimensional theory of a massless scalar field in the presence of a static $\delta$-like potential. First of all, we demonstrated that this problem is exactly solvable in $\mathrm{GF_1}$ theory and presented this solution explicitly.

In a local quantum theory after the frequency Fourier transform, the problem reduces to solving a time-independent Schr\"odinger-type equation \eqref{VEQ}. The non-local ghost-free generalization produces a similar equation \eqref{GF}, and one can easily adapt the solution of the latter equation, presented in this letter, to a more general case when the field is massive, with mass $m$, and the number of spatial dimensions is greater than 1. To include a non-vanishing scalar mass it is sufficient to modify the parameter $E$, which enters \eqref{DE}, according to $E=\omega^2-m^2$. If the number of spatial dimensions is greater than 1, and the potential $V$ does not depend on the transverse coordinates $\vec{y}_{\perp}$, then one can write mode-solutions in the form
\be
\varphi_{\omega,\vec{k}_{\perp}}=\varphi_{\omega}\exp(i\vec{k}_{\perp}\!\cdot\vec{y}_{\perp})\, .
\ee
By performing the Fourier transform with respect to the transverse coordinates $\vec{y}_{\perp}$ one again obtains the equation \eqref{GF} with the only change $E=\omega^2-m^2-\vec{k}_\perp^2$.

When we started this project, our naive expectations were that in the ghost-free theory the non-locality effectively smears the sharp $\delta$-like potential, and, as a result, the scattering coefficients would be only slightly modified. Calculations based on the derived exact solution in the $\mathrm{GF_1}$ theory demonstrated many interesting unexpected features.

For the $\delta$-potential barrier there exists a dimensionless value of the frequency, $\alpha_\star$, at which the transmission coefficient vanishes identically, which is a property that does not occur in the local case. For the $\delta$-potential well ($\lambda<0$) in the non-local case there also exist quasi-normal modes corresponding to $E<0$. The quasi-normal mode with purely imaginary frequency is unique, as in the local theory. However, if the strength of the potential well is above some critical value, $|\Lambda| > 2.282365$, there appear extra quasi-normal modes with complex frequencies.

In this letter, we presented results for a special choice of the ghost-free theory. Nonetheless, we expect that similar results can be obtained for more complicated cases, for example when the non-local form-factor  is of the form $\sim \exp{[(-\Box \ell^2)^N]}$. However, the technical details become more involved and exact solutions may not be readily obtained for these generalizations.

Lastly, let us point out that the method developed in this letter can be generalized to the case of multiple $\delta$-like potentials. Our preliminary analysis shows that these problems are also exactly solvable. One of the possible natural interesting applications is the calculation of the Casimir force and energy in ghost-free quantum theory. We hope to present our results on this subject in a separate publication.

\section*{Acknowledgments}

J.B.\ is grateful for a Vanier Canada Graduate Scholarship administered by the Natural Sciences and Engineering Research Council of Canada as well as for the Golden Bell Jar Graduate Scholarship in Physics by the University of Alberta.
V.F.\ and A.Z.\ thank the Natural Sciences and Engineering Research Council of Canada and the Killam Trust for their financial support.
\vfill

\bibliography{Ghost_references}{}

\end{document}